\documentclass{PoS}

\title{Shadow of black holes at local and cosmological distances}

\ShortTitle{Shadow of black holes}

\author{\speaker{G.S. Bisnovatyi-Kogan}\\
        Space Research Institute,  Russian Academy of Sciences,\\ Profsoyuznaya 84/32, Moscow 117997, Russia\\ and \\National Research Nuclear University MEPhI \\ (Moscow Engineering Physics Institute), Kashirskoe Shosse 31, Moscow 115409, Russia\\and\\
        Moscow Institute of Physics and Technology MIPT, Dolgoprudny, Moscow reg., Russia\\
        E-mail: \email{gkogan@iki.rssi.ru}}

\author{O.Yu. Tsupko\\
        Space Research Institute,  Russian Academy of Sciences,\\ Profsoyuznaya 84/32, Moscow 117997, Russia\\
       E-mail: \email{tsupko@iki.rssi.ru}}

\author{V. Perlick\\
       ZARM, University of Bremen, Germany
\\
       E-mail: \email{perlick@zarm.uni-bremen.de}}

\abstract{
A brief illustrative discussion of the shadows of black holes at local and cosmological distances is presented. Starting from definition of the term and discussion of recent observations, we then investigate shadows at large, cosmological distances. On a cosmological scale, the size of shadow observed by comoving observer is expected to be affected by cosmic expansion. Exact analytical solution for the shadow angular size of Schwarzschild black hole in de Sitter universe was found. Additionally, an approximate method was presented, based on using angular size redshift relation. This approach is appropriate for general case of any multicomponent universe (with matter, radiation and dark energy). It was shown, that supermassive black holes at cosmological distances in universe with matter may give the shadow size comparable with the shadow size in M87, and in the center of our Galaxy.
}

\FullConference{Multifrequency Behaviour of High Energy Cosmic Sources - XIII - MULTIF2019\\
		3-8 June 2019\\
		Palermo, Italy}

\begin{document}

\section{What is a black hole shadow}

\begin{figure}
	\center
{\includegraphics[width=5in]{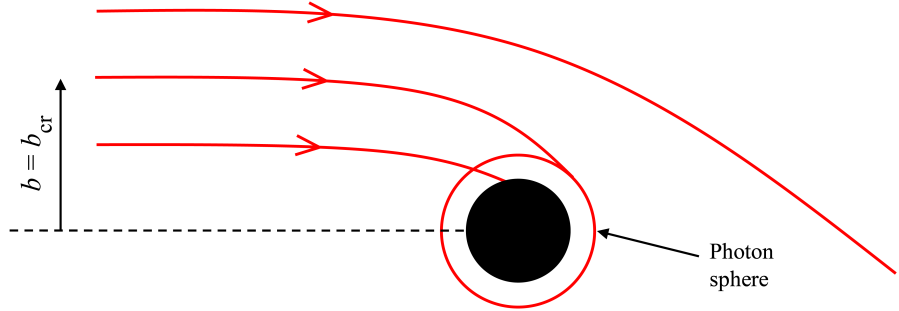}
	\caption{Trajectory of a photon in vacuum is determined by its impact parameter. Photon sphere plays crucial role in formation of the shadow, from \cite{bkt17}. \label{fig1}
}}
	    \end{figure}

A black hole (BH) absorbs all photons in its vicinity, so, if there are no other sources on the line of sight, the observer is finding a dark spot in the direction of BH, which is called as BH shadow. Angular size of the shadow of a Schwarzschild BH in vacuum is defined by the formula \cite{Synge1966}
\begin{equation}
\sin ^2 \alpha_{\mathrm{sh}} \, = \,
\frac{27m^2 (1-2m/r_O)}{r_O^2}.
\label{eq1}
\end{equation}
Here $r_O$ is the radial coordinate of observer, the Schwarzschild radius $R_S=2m$, radius of the photon sphere is $r_{ph}=3m$, $m=GM/c^2$ is mass parameter. For a distant observer, $r_O \gg m$, the shadow angular radius reduces to
$\alpha_{\mathrm{sh}} \simeq 3\sqrt{3}\frac{m}{r_O}$. The schematic picture of the photon motion in the vicinity of BH for different impact parameters $b$ is given in Fig.\ref{fig1}.

\begin{figure}
	\center
{\includegraphics[width=5in]{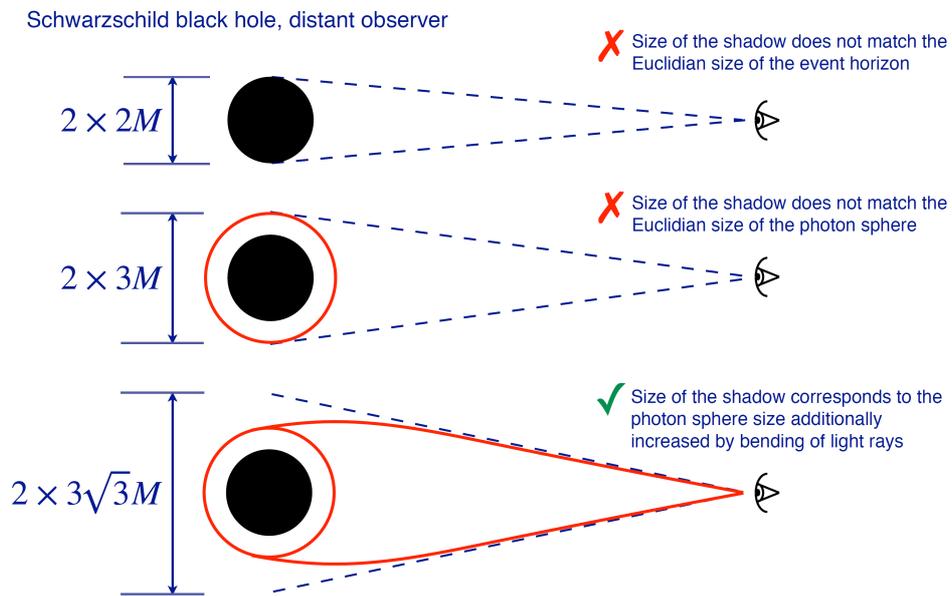}
	\caption{Some misconceptions about the BH shadow, for a distant observer. Image by O. Tsupko. \label{fig4}
}}
	    \end{figure}

\begin{figure}
	\center
{\includegraphics[width=5in]{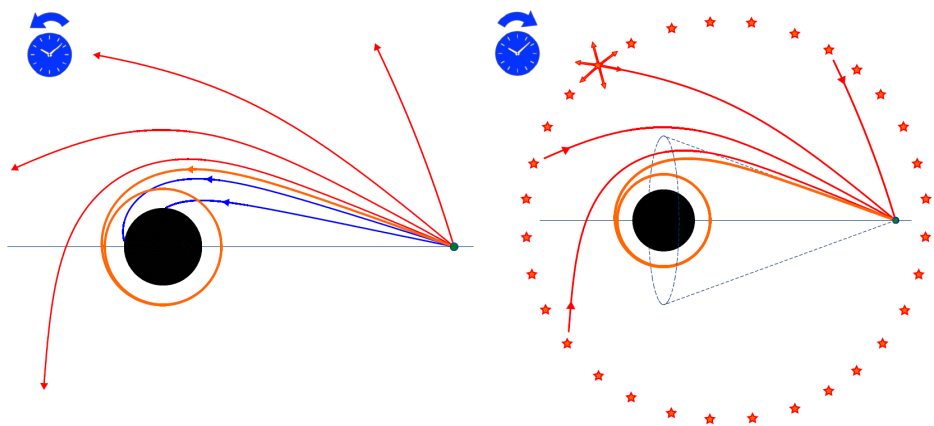}
	\caption{The shadow is defined as a region of the observer's sky that is left dark if there are light sources distributed everywhere but not between the observer and the black hole.
{\bf Left}: observer emits light rays to the past. Some rays are deflected, other rays are absorbed by black hole. Boundary between these two classes of light rays is light rays which go to the photon sphere
{\bf Right}: there are many sources around, every source emits light rays to all directions. But shown angular cone will be empty for observer. This is the shadow. Image by O. Tsupko. \label{fig5}
}}
	    \end{figure}

\begin{figure}
	\center
{\includegraphics[width=5in]{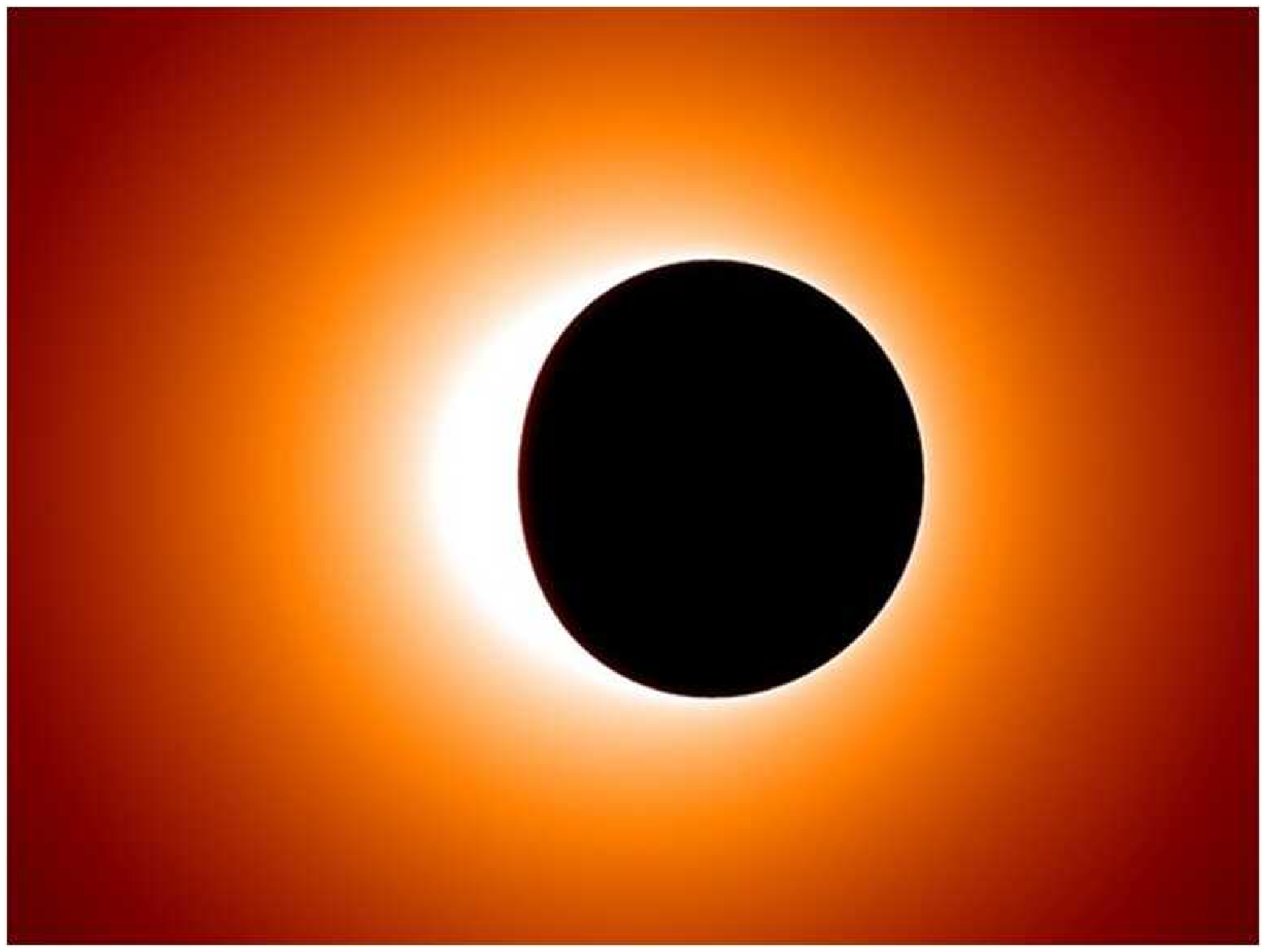}

\bigskip

	\caption{General relativistic ray tracing simulations of the black hole. For rotating black hole the shadow is not symmetrical although metric is axially symmetric. Shadow is deformed and oblate. Image by M. Moscibrodzka and H. Falcke \cite{web1}. \label{fig6}
}}
	   \end{figure}

\noindent The angular size of a shadow is connected with existence of the horizon, but it exceeds the angular size of the horizon due to falling into BH of all photons inside the photon sphere (Fig.\ref{fig1}), and due to curved trajectories of photons coming from the photon sphere itself. This phenomena is illustrated in Figs.\ref{fig4},\ref{fig5}. \noindent Analytical calculation of the shadow size and shape in vacuum for Kerr metric was first done by J.M. Bardeen \cite{Bardeen1973}, for the observer far away from BH, see Fig.\ref{fig6}, \cite{perlick2014}, \cite{Tsupko2017}, \cite{cunha2018}.

\section{Observation of the shadow. Shadow of supermassive black holes}

\begin{figure}
	\center
{\includegraphics[width=5in]{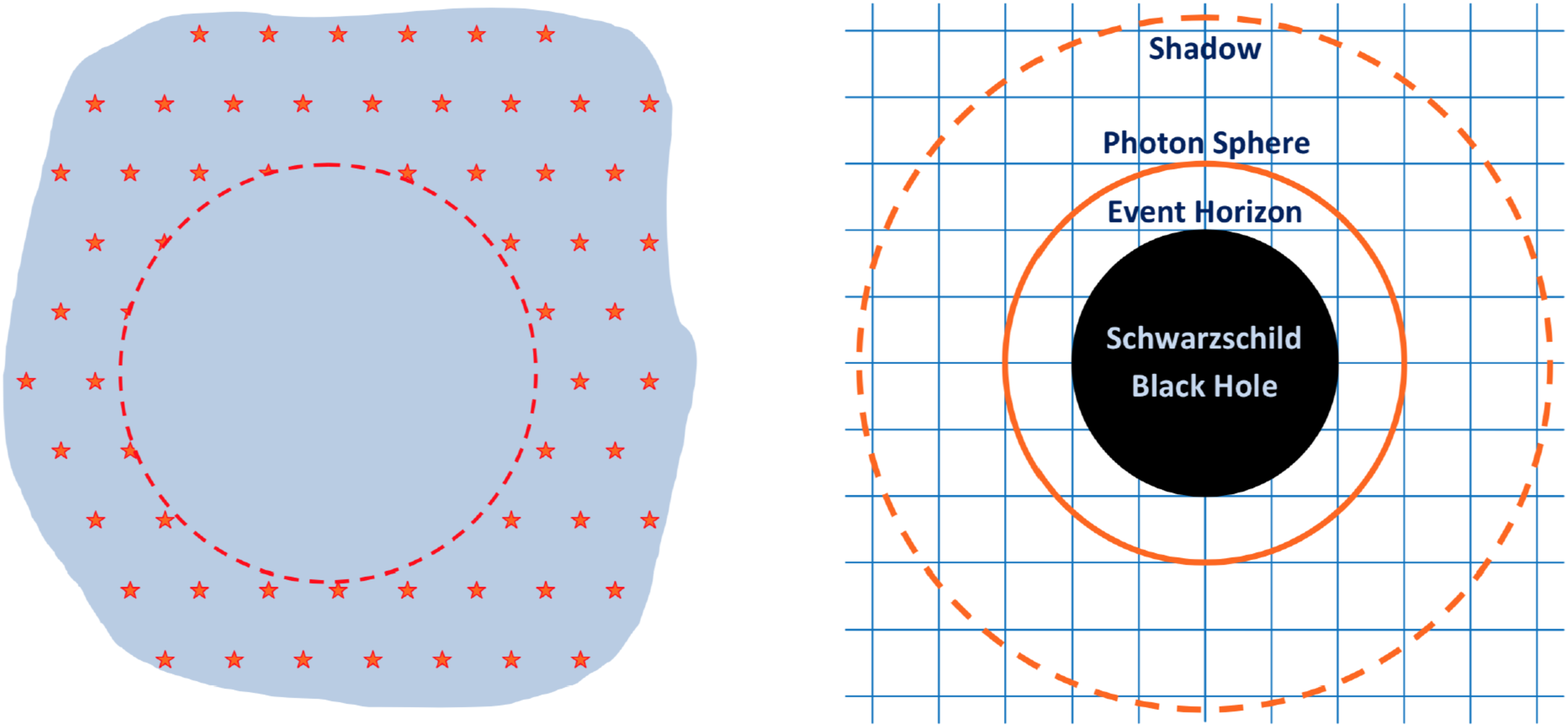}
	\caption{\textbf{Left}: Dark spot (shadow) at the background of bright sources.  \textbf{Right}: Comparison of BH shadow size with Euclidean sizes of the photon sphere and event horizon, for a distant observer. Image by O. Tsupko \label{fig7}
}}
	    \end{figure}

\begin{figure}
	\center
{\includegraphics[width=2.5in]{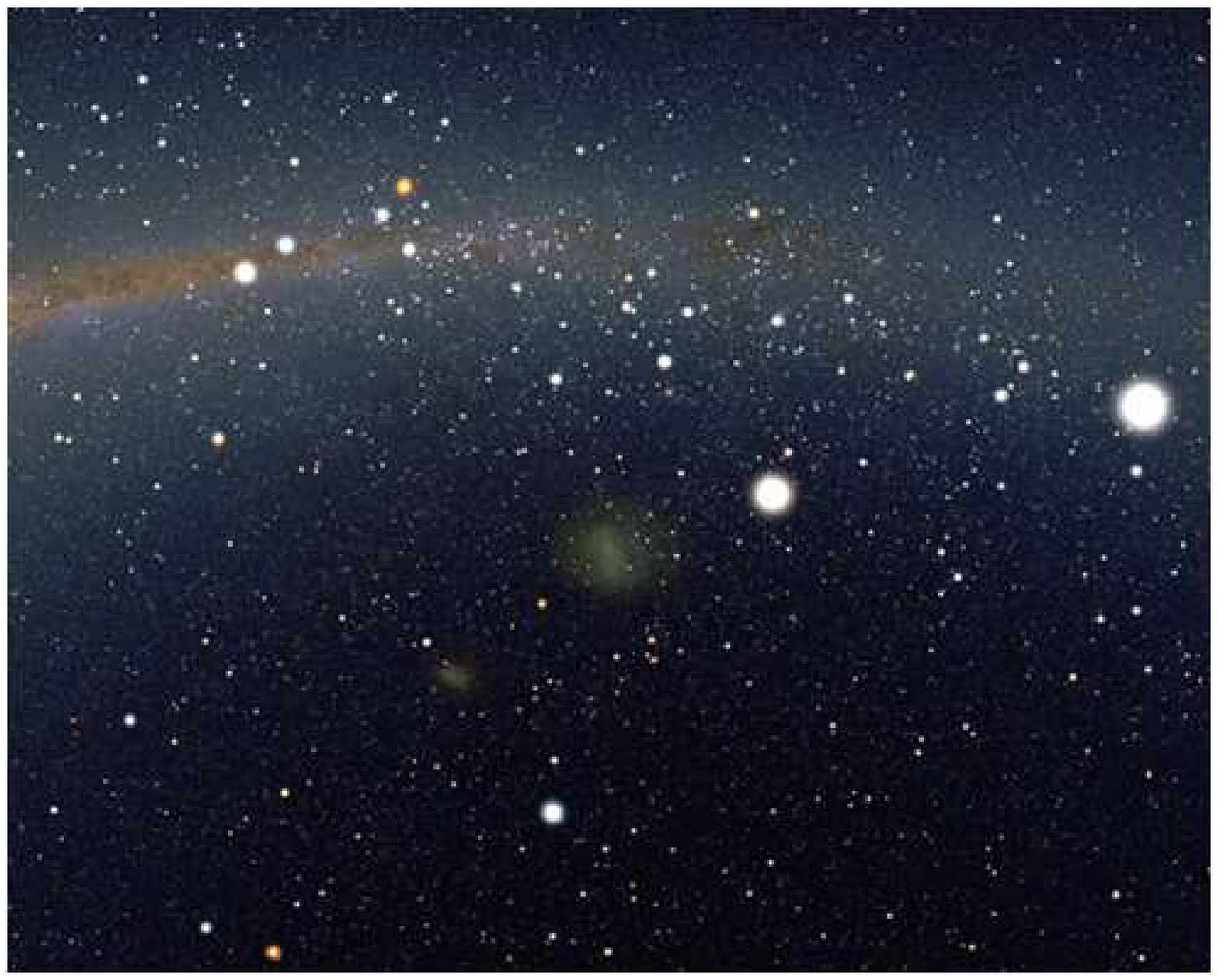}
 \includegraphics[width=2.5in]{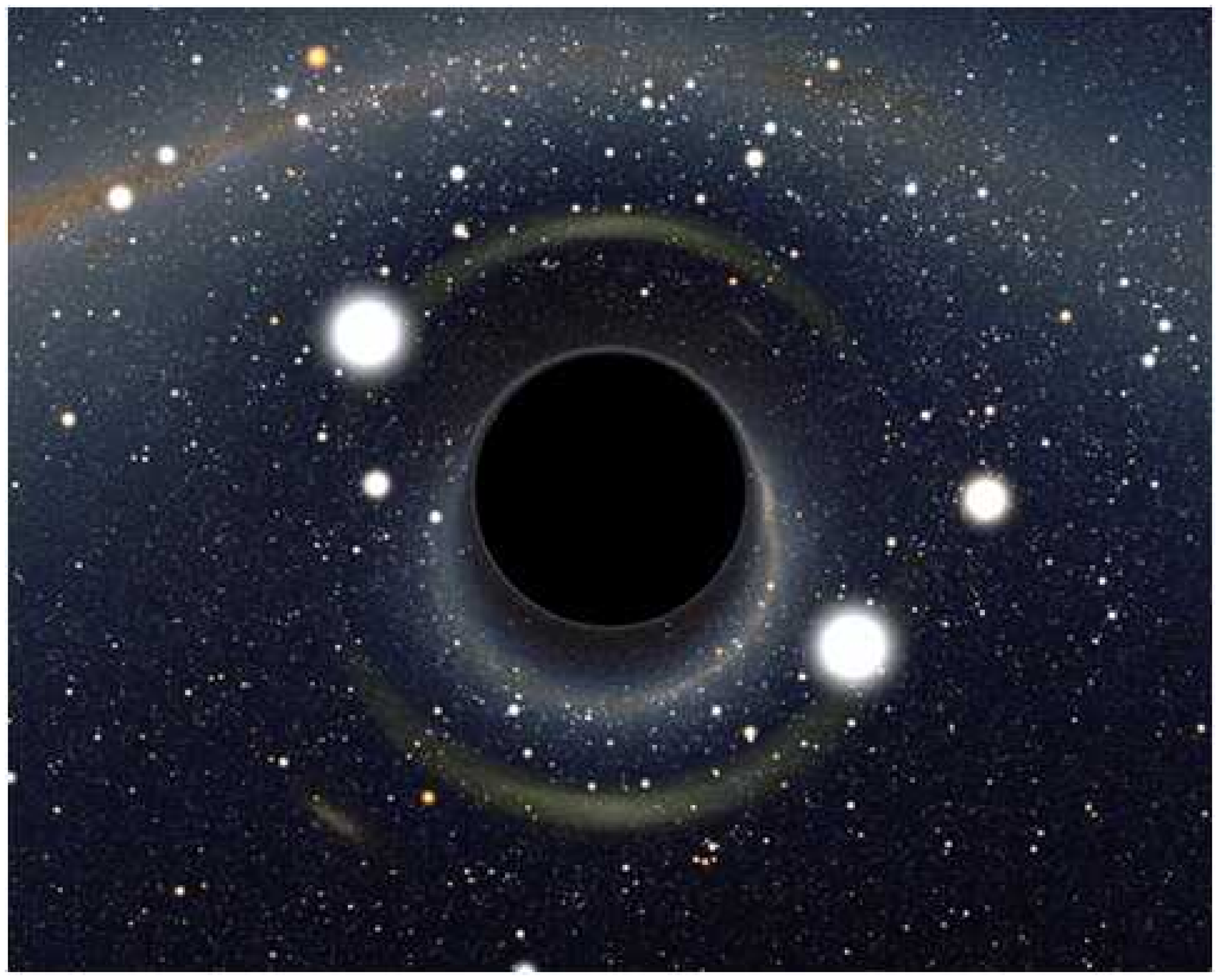}
	\caption{Simulation of starlight lensed by a Schwarzschild black hole.
{\bf Left.} Region of the sky around the Large Magellanic Cloud in the southern hemisphere. {\bf Right.} Same region, distorted by the presence of a foreground black hole. Each star position has changed. Strong distortions are visible for extended objects such as the Large Magellanic Cloud. Any star now has a ghost image. Few ones  have a large increase in luminosity (as well as their ghost image) due to gravitational  lensing of stars which are close to being exactly behind the black hole. Credit: A. Riazuelo, IAP/UPMC/CNRS \cite{LMC}. \label{fig8}
}}
	    \end{figure}

\begin{figure}
	\center
{\includegraphics[width=5in]{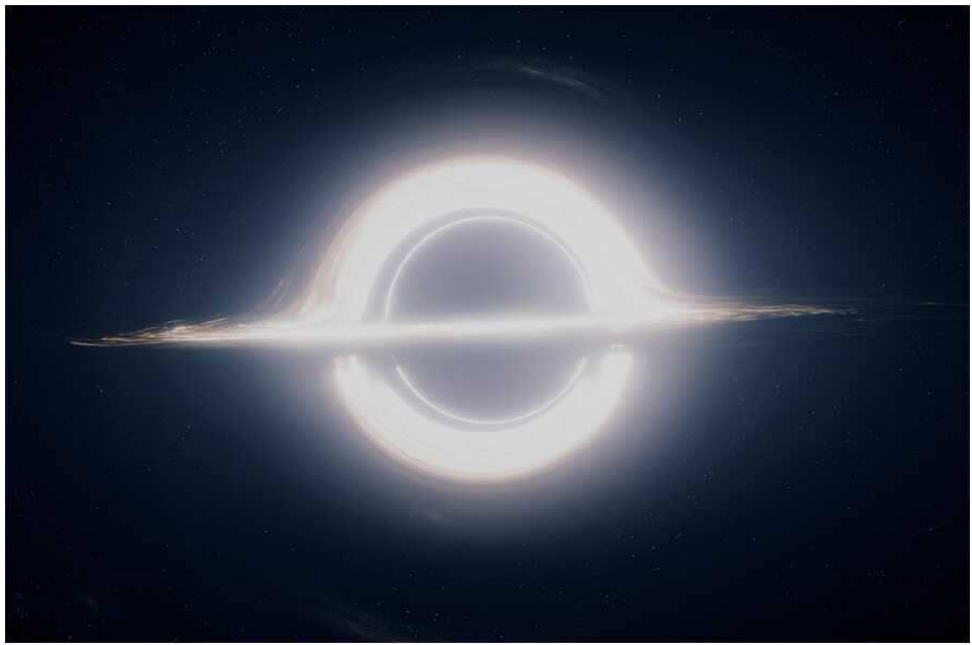}

\bigskip

	\caption{Black hole shadow + lensed image of accretion disk. Movie Interstellar, Warner Bros. Entertainment Inc.. from \cite{LMC}, see also \cite{james2015}. \label{fig9}
}}
	   \end{figure}

\begin{figure}
	\center
{\includegraphics[width=5in]{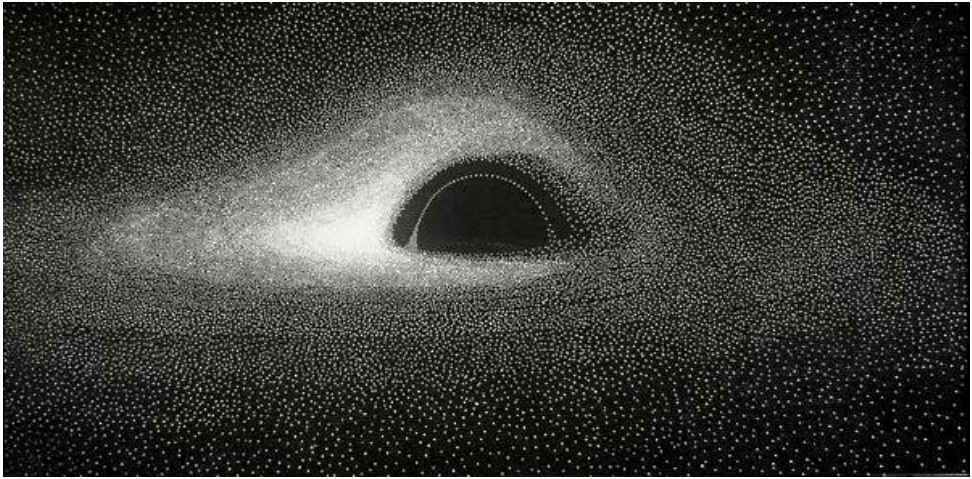}
	\caption{Simulations of a shadow of BH, surrounded by accretion disk, from \cite{luminet1979,web6}. Luminet calculated in 1979 using the IBM 7040 mainframe, an early transistor computer with punch card inputs. Using numerical data from the computer, he drew directly on negative image paper with black India ink, placing dots more densely where the simulation showed more light.
 \label{fig10}
}}
	    \end{figure}

Supermassive BH are present in the center of galaxies, for example, in the center of our Galaxy.
A distant observer should 'see' this black hole as a dark disk in the sky at the background of bright sources which is known as the 'shadow' \cite{fma2000} (see Fig.\ref{fig7}). The lensing effect acting in the vicinity of BH shadow
leads to strong distortion of the visible picture, compared to the real one, appearance twin images etc., see Fig.\ref{fig8}. Simulations of a shadow of BH, surrounded by accretion disk, visible from different angles, are given in Figs.\ref{fig9},\ref{fig10}. For the black hole at the center of our galaxy, size of the shadow is about 53 $\mu as$ (size of grapefruit on the Moon). Observational project using (sub)millimeter VLBI observations with telescopes distributed over the Earth is Event Horizon Telescope (http://eventhorizontelescope.org), see also BlackHoleCam (http://blackholecam.org).

\begin{figure}
	\center
{\includegraphics[width=5in]{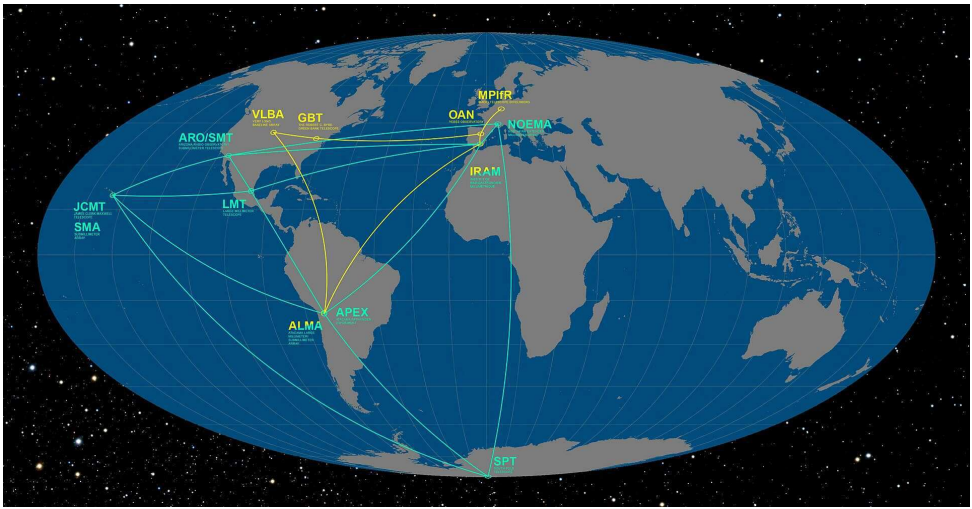}

\bigskip

	\caption{Event Horizon Telescope. Credit: ESO/O. Furtak \cite{m87a,web2}. \label{fig11}
}}
	    \end{figure}

\begin{figure}
	\center
{\includegraphics[width=5in]{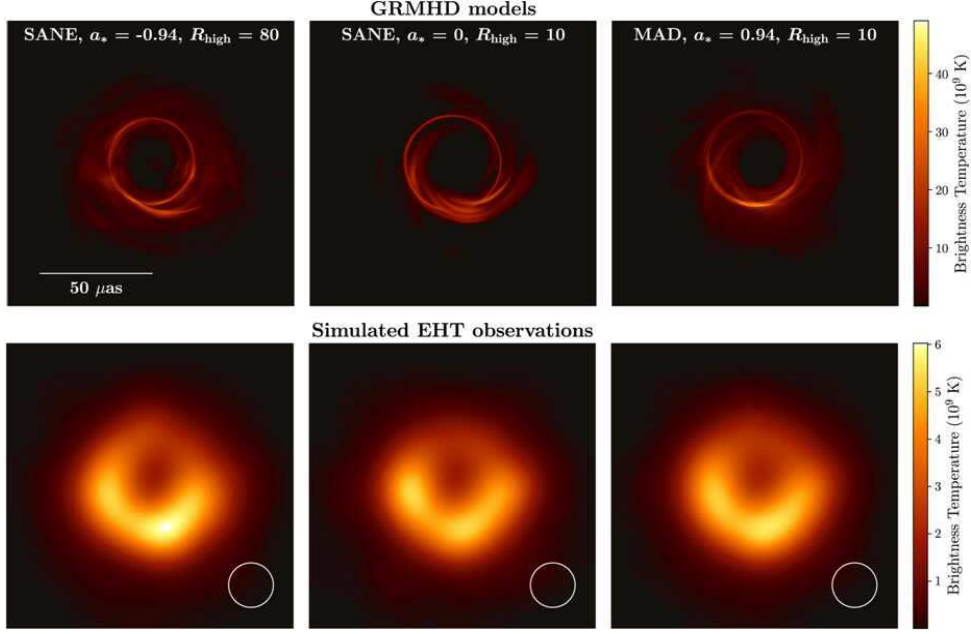}

\bigskip

	\caption{Simulations of a shadow of BH, observed by EHT, from \cite{m87,web3}. \label{fig12}
}}
	    \end{figure}

\noindent The Event Horizon Telescope (EHT) is a planet-scale array of eight ground-based telescopes. Project use (sub)millimeter
VLBI observations with telescopes distributed over the Earth, see Fig.\ref{fig11}.
The data put Einstein's theory to what could be its most rigorous test yet; the size and shape of the observed
black hole can all be predicted by the General Relativity equations, which posits it to be roughly circular, see Fig.\ref{fig12}.
The simulations (top row) are shown above alongside the simulated image of the black hole itself (bottom row).

\noindent Estimations before BH shadow observations gave

 Sgr A*:  $ M = 4.3 \times 10^6 M_{\odot},\quad D = 8$ kpc, angular diameter of the shadow is about 53 $\mu$as.

\noindent  M87 (thousand times more massive and thousand times more distant):

1) $M = 6 \times 10^9 M_{\odot},\quad D = 16$ Mpc, angular diameter of the shadow is about 38 $\mu$as.

2) $M = 3.5 \times 10^9 M_{\odot},\quad D = 18$ Mpc, angular diameter of the shadow is about 20 $\mu$as.

\begin{figure}
	\center
{\includegraphics[width=5in]{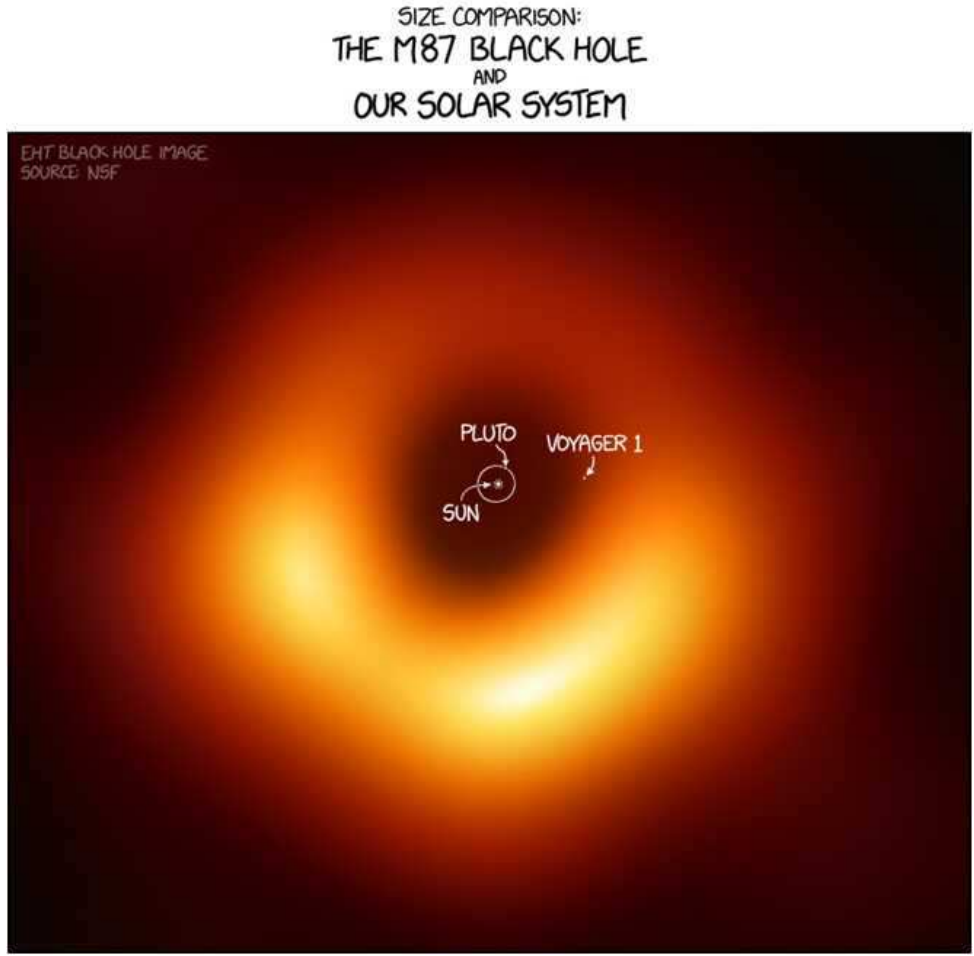}

\bigskip

	\caption{First M87 Event Horizon Telescope Results.
The Shadow of the Supermassive Black Hole: how big the M87 black hole is compared to the Solar system. Image by Randall Munroe, from \cite{web4}. \label{fig13}
}}
	    \end{figure}

\noindent Current observations in M87 correspond to the
Ring Diameter = $42 \pm 3\,\mu as$. It gives the mass $(6.5 \pm 0.7)\times 10^9\, M_\odot$, for the adopted distance of $16.8 \pm 0.8$ Mpc.
The observed shadow of the supermassive black hole in M87, in comparison with the objects of the Solar system, is given in Fig.\ref{fig13}. The model of the observed shadow formation is also investigated in \cite{wald2019}.

\section{Influence of the cosmological expansion on the shadow size. Exact solution for Kottler metric}

In the empty universe with a cosmological constant the size of a shadow of the black hole, for a comoving observer, was calculated analytically in \cite{ptbk2018}.
The Kottler metric is the unique spherically symmetric solution to
Einstein's vacuum field equation with a cosmological constant. In its standard static form it
reads
\begin{equation}\label{eq:Kottlerstat}
g_{\mu \nu} dx^{\mu} dx^{\nu}
=  -f(r) c^2 dt^2 + \frac{dr^2}{f(r)} + r^2 d \Omega ^2
\end{equation}
where
\begin{equation}\label{eq:fOmega}
f(r) = 1 - \frac{2m}{r} - \frac{\Lambda}{3} r^2 \, , \quad
d \Omega ^2 = \mathrm{sin} ^2 \vartheta \, d \varphi ^2 + d \vartheta ^2 \, ,
\end{equation}
$m$ is the mass parameter,
\begin{equation}\label{eq:m}
m = \frac{GM}{c^2} ,
\end{equation}
where $M$ is a mass of the central object, and $\Lambda$ is a
cosmological constant.
  In the case of  BH with the Schwarzschild metric, the angular size of the
shadow, seen by a static observer (\ref{eq1})  can be written as \cite{Synge1966}

\begin{equation}
\label{eq:theta4}
\mathrm{sin} ^2 \theta_{stat} \, = \,
\frac{(1- \frac{2m}{r_O}) b_{cr}^2}{
 r_O^2} \, ,
\end{equation}
where $b_{cr}$ is the critical value of the impact parameter $b=cL/E$ ($E$ is the energy, $L$ is the angular momentum of the photon), corresponding to photons in the unstable circular orbits, filling the photon sphere. For the Schwarzschild metric, the radius of the photon sphere equals $3m$, and $b_{cr}=3\sqrt{3} \, m$. For large distances the angular size of the shadow is written as

\begin{equation}
\label{bigD}
\sin^2\theta_{stat} \approx \frac{b_{cr}^2}{r_O^2}, \qquad r_O\gg m.
\end{equation}
Black hole angular shadow size, as seen by static observer in Kottler (Schwarzschild-de Sitter) space-time (\ref{eq:Kottlerstat}), is written as \cite{sh1999}
\begin{equation}
\label{eqketl}
\mathrm{sin} ^2 \theta_{stat} \, = \,
\frac{(1- \frac{2m}{r_O}- \frac{\Lambda}{3} r_O^2) b_{cr}^2}{
 r_O^2} \, ,
\end{equation}
with the critical impact parameter
\begin{equation}
\label{eqketl1}
b_{cr}=\frac{3\sqrt{3} \, m}{(1-9\Lambda m^2)^{1/2}}.
\end{equation}
For the Kottler spacetime the determination of the angular size of the shadow at large distances does not reduce to the calculation of the critical value of the impact parameter:
\begin{equation}
\sin^2\theta_{stat} \not\approx \frac{b_{cr}^2}{r_O^2}, \qquad r_O\gg m.
\end{equation}

\begin{figure}
	\center
{\includegraphics[width=5in]{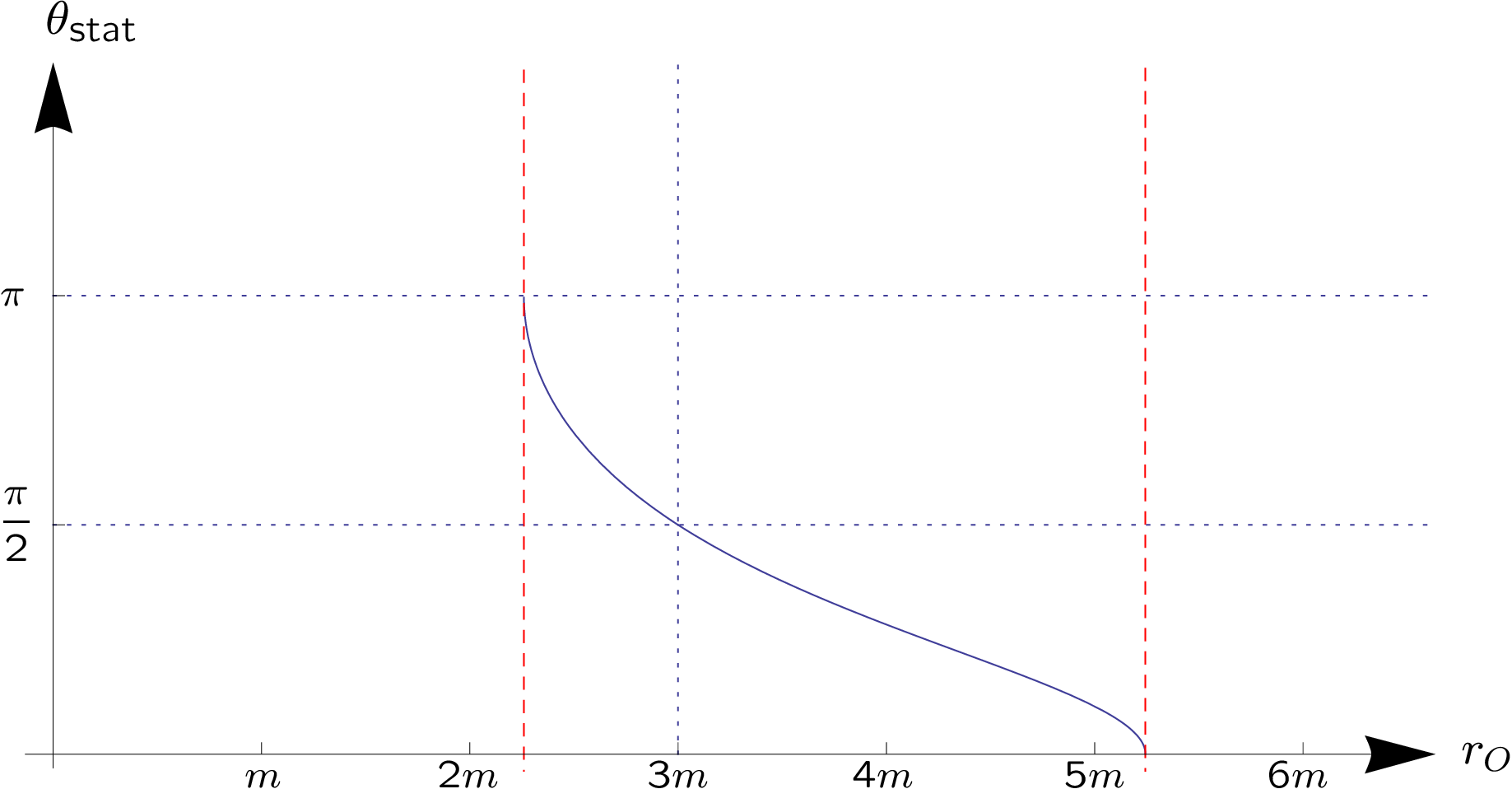}
%
%
	\caption{Angular radius $\theta _{\mathrm{stat}}$ of the shadow
		plotted against the observer position $r_O$. The picture is
		for $\sqrt{\Lambda /3} = H_0/c= 0.15 \, m^{-1}$. The dashed (red) lines
		mark the horizons at $r=r_{\mathrm{H1}}$ and $r=r_{\mathrm{H2}}$, from \cite{ptbk2018}. \label{fig14}
}}
	    \end{figure}

The shadow angular size seen by a static observer is plotted in Fig.\ref{fig14}, it goes to zero on the outer horizon $r_{H2}$ at large $r_O$, connected with $\Lambda$, where for very small realistic $\Lambda$ we have $r_O\approx \sqrt{\frac{3}{\Lambda}}$, $r_{H1}$ is the inner horizon, connected with a Schwarzschild singularity.

\begin{figure}
	\center
{\includegraphics[width=5in]{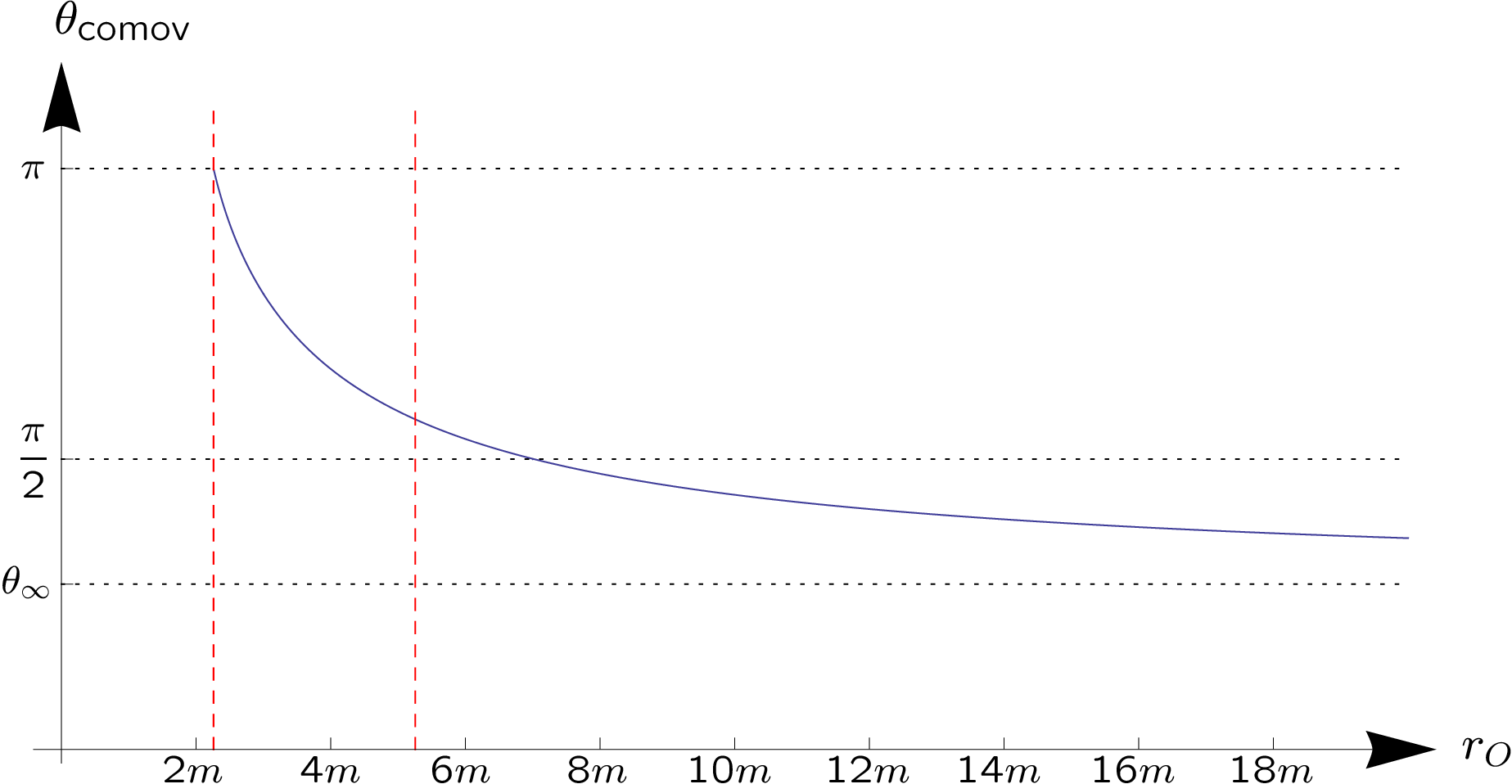}
%
%
	\caption{ Angular radius $\theta _{\mathrm{comov}}$ of the shadow
		plotted against the observer position $r_O$. As before, we have
		chosen $\sqrt{\Lambda /3} = H_0/c= 0.15 \, m^{-1}$ and the dashed (red) lines
		mark the horizons,  from \cite{ptbk2018}. \label{fig15}
}}
	    \end{figure}

\noindent The shadow angular radius as seen by comoving observer in plotted in Fig.\ref{fig15}, and is defined by the expression \cite{ptbk2018}
\begin{equation}
\mathrm{sin} \, \theta _{\mathrm{comov}} =
\frac{\sqrt{27} \, m}{r_O} \sqrt{1-\frac{2m}{r_O}}
\sqrt{ 1 - \frac{27H_0^2m^2}{c^2}} \, \mp \, \frac{\sqrt{27} \, m \, H_0}{c}
\sqrt{ 1 - \frac{27 m^2}{r_O^2} \Big( 1 - \frac{2m}{r_O} \Big)}
\, ,
\label{eq:comovsh1}
\end{equation}

\noindent where $H_0=c\sqrt{\frac{\Lambda}{3}}$.
This equation makes sense for all momentary observer positions
$r_O$ with $r_{H1} < r_O < \infty$. For the inner horizon, we have to use the upper sign in
(\ref{eq:comovsh1}), so that

\vspace{-0.15cm}

\begin{equation}\label{eq:inner}
\mathrm{sin} \,  \theta _{\mathrm{comov}} \to 0 \quad
\mathrm{for} \quad r_O \to r_{H1} \, .
\end{equation}

\vspace{0.15cm}

\noindent and the angle $\theta _{\mathrm{comov}}$ itself goes to $\pi$. For the outer horizon, however, we have to use the lower sign
in (\ref{eq:comovsh1}), when we have an asymptotic behaviour as

\begin{equation}\label{eq:inf}
\mathrm{sin} \,  \theta _{\mathrm{comov}} \to
\sqrt{27} \, \frac{H_0m}{c}
\quad \mathrm{for} \quad r_O \to \infty \, .
\end{equation}
The change of the sign in (\ref{eq:comovsh1}) should be done at zero value of the second term, what happens at $r_O=3m$.
When the comoving observer starts at the inner horizon, the shadow covers
the entire sky, $\theta _{\mathrm{comov}} = \pi$. On his way out to
infinity, the shadow monotonically shrinks to a \emph{finite} value given by
(\ref{eq:inf}), see Fig. \ref{fig15} from \cite{ptbk2018}. For a supermassive black hole of
$M=10^{10}\, M_\odot$ the angular size goes to $\theta_{\mathrm{comov}} \approx 0.1\,\,\mu$as in the limit $r_O \rightarrow \infty$, accepting the present value of $H_0^{-1}=5\cdot 10^{17}$ s, corresponding to the modern notation $H_m=70\,\,$km/s/Mpc.

\section{Approximate solution for arbitrary expanding universe}

Difficulties to derive exact analytical solution for general case are connected with absence of an exact analytical solution for geodesics, in the expanding universe, containing both, matter and $\Lambda$, in presence of a central BH.
We have obtained an approximate analytic solution for a size of the shadow of a  supermassive BH in the expanding universe, at cosmological distance, with $r_O \gg m$ \cite{bkt18}.
We have used a well-known effect of increase of apparent angular size of the object if observed by comoving observer in the expanding universe
\cite{Mattig1958, Zeldovich1964, dz1965}.
In the modern literature, it is described in terms of so called angular size redshift relation which relates apparent angular size of the object of a given physical size, and its redshift $z$ \cite{web5}.

According to definition of angular diameter distance $D_A$, we write:
\begin{equation}
D_A = \frac{L}{\Delta \theta} \, ,
\end{equation}
where $L$ is the proper diameter of the object, and $\Delta \theta$ is the observed angular diameter. For given universe, angular diameter distance is known function of the redshift $z$ (we restrict ourselves by the flat case):
\begin{equation}
D_A(z) = \frac{c}{(1+z) H_0} Int(z) \, ,
\end{equation}
where
\begin{equation}
Int(z) = \int \limits_0^z \left( \Omega_{m0} (1+\tilde{z})^3 + \Omega_{r0} (1+\tilde{z})^4 + \Omega_{\Lambda 0} \right)^{-1/2} d\tilde{z} \, .
\label{int}
\end{equation}
Here $H_0$ is the present day value of the Hubble parameter $H(t)$, and $\Omega_{m0}$, $\Omega_{r0}$, $\Omega_{\Lambda 0}$ are the present day values of density parameters for matter, radiation and dark energy correspondingly.

Angular size redshift relation cannot be directly applied for exact analytical calculation of shadow size. Formula for angular size redshift relation is based on assumption that light rays propagates in expanding FRW metric without other additional sources of gravity. For shadow it is not the case.
Therefore, exact analytical solution for the shadow size in expanding universe cannot be written using angular size redshift relation. Nevertheless, it turns out that in some approximations it becomes possible to find the approximate solution with help of this relation.

We propose an approximate method of calculation of the shadow in expanding Friedmann universe as seen by comoving observer which is appropriate for a general case (with matter, radiation and dark energy).
In realistic situations:

1. The observer is very far from the BH, at distances much larger than BH horizon;

2. The expansion is slow enough to be significant only on very large scales.

\noindent Therefore we can neglect

1. Influence of the expansion on a particle motion near the BH,

2. BH gravity during a long light travel to the observer if he is distant enough.

    \noindent  For the region near BH there are formulas for shadow (without expansion), and we can calculate effective linear size of the shadow in this region.
      In region far from BH, there are formulas for angular size redshift relation in FRW metric (without BH gravity).
      Therefore we can calculate the shadow in the following way: find some 'effective' linear size of the shadow near BH and then substitute it into angular size redshift relation. This will give the correct approximate solution for general case.

\bigskip

\begin{figure}
	\center
{\includegraphics[width=5in]{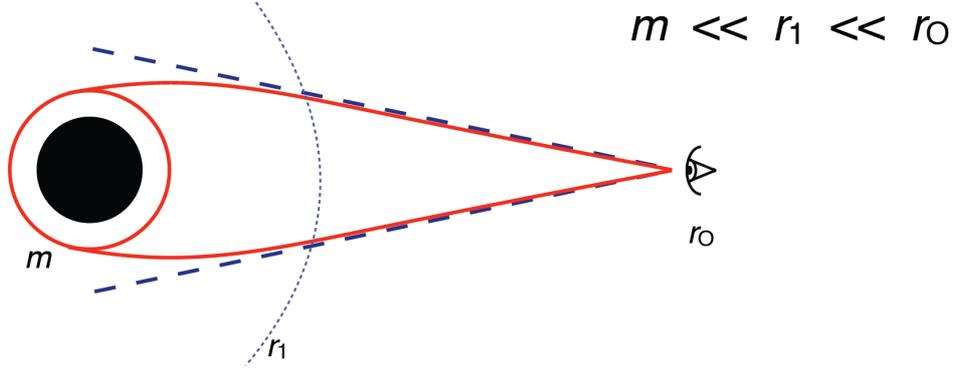}

\bigskip

	\caption{The dash line is situated at $r=r_1$, which is large enough in comparison with the BH horizon, to neglect the BH gravity,  it  is small enough in comparison with observer's coordinate $r_o$, to neglect the cosmic expansion. Near $r_1$ the space-time is almost flat. Image by O. Tsupko. \label{fig16}
}}
	    \end{figure}

Let us choose a radial coordinate $r_1$, such as (see Fig.\ref{fig16})
\begin{equation}
m\ll r_1 \ll r_O,\quad  2m\,\, {\mbox{ is a radius of the horizon of a BH}}.
\end{equation}
Assume that coordinate $r_1$ is large enough in comparison with the BH horizon, to neglect the BH gravity. At the same time $r_1$ is small enough in comparison with observer's coordinate to neglect the cosmic expansion. Around $r_1$ the space-time is almost flat.
In the first region $(r < r_1)$ we neglect expansion and calculate the effective linear size of the shadow by standard formulas.
In the second region $(r > r_1)$ the cosmic expansion predominates, and we neglect BH gravity. Here we use angular size redshift relation. A crucial thing to know for is an effective linear size of the shadow.
We have found that effective linear radius equals to $3\sqrt{3}\, m$.

\begin{figure}
	\center
		\includegraphics[width=5in]{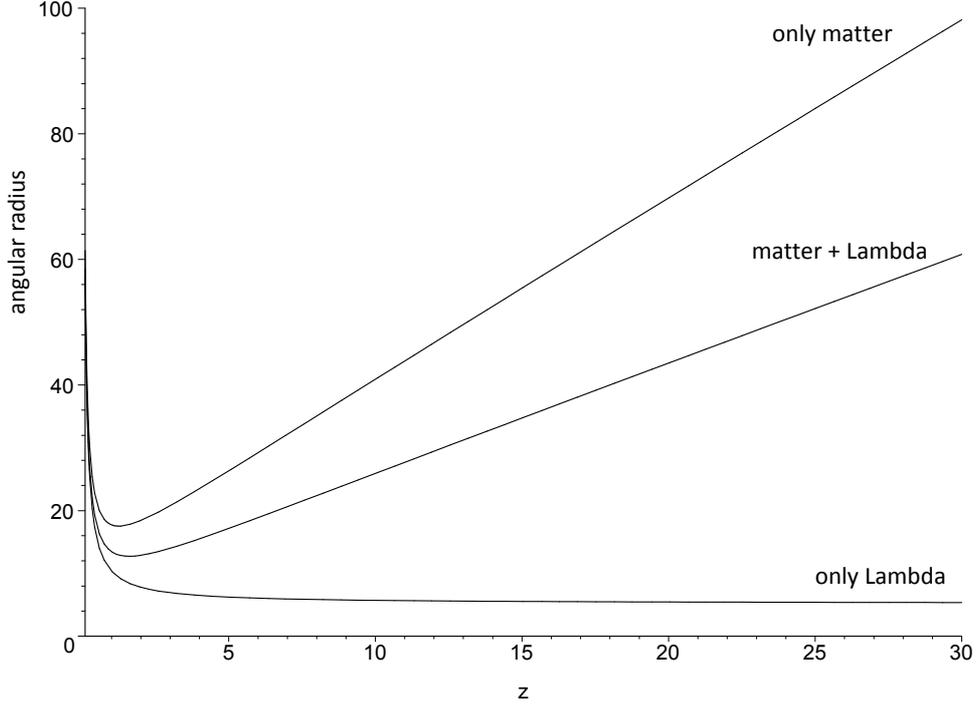}
	\caption{Dependence of angular radius $\alpha_{\mathrm{sh}}$ on $z$ for different spacetimes: de Sitter model without matter ($\Omega_{m0}=0$, $\Omega_{\Lambda 0}=1$), Einstein--de-Sitter spacetime filled by matter only ($\Omega_{m0}=1$, $\Omega_{\Lambda 0}=0$) and mixed case with matter and dark energy presence ($\Omega_{m0}=0.3$, $\Omega_{\Lambda 0}=0.7$). Angular radius is presented in units of $mH_0/c$, from \cite{bkt18}. \label{fig17}}
\end{figure}

Finally, the approximate formula for the visible angular radius of the shadow $\alpha_{sh}$ is written as

\begin{equation} \label{main-result}
\alpha_{\mathrm{sh}}(z) = \frac{3\sqrt{3}m}{D_A(z)} = 3\sqrt{3}m \, \frac{H_0}{c} \, \frac{1+z}{Int(z)} \, .
\label{shadz}
\end{equation}
with $Int(z)$ from (\ref{int}). This formula allows to calculate the size of shadow as function of black hole redshift in universe with given parameters $H_0$, $\Omega_{m0}$, $\Omega_{r0}$, $\Omega_{\Lambda 0}$.
It agrees with the exact analytical solution for pure Lambda case with corresponding approximations,
and gives  correct asymptotic behaviour for small $z$.
The dependence $\alpha_{sh}(z)$, according to (\ref{int}),(\ref{shadz}) is plotted in Fig.\ref{fig17}.

\begin{figure}
	\center
{\includegraphics[width=5in]{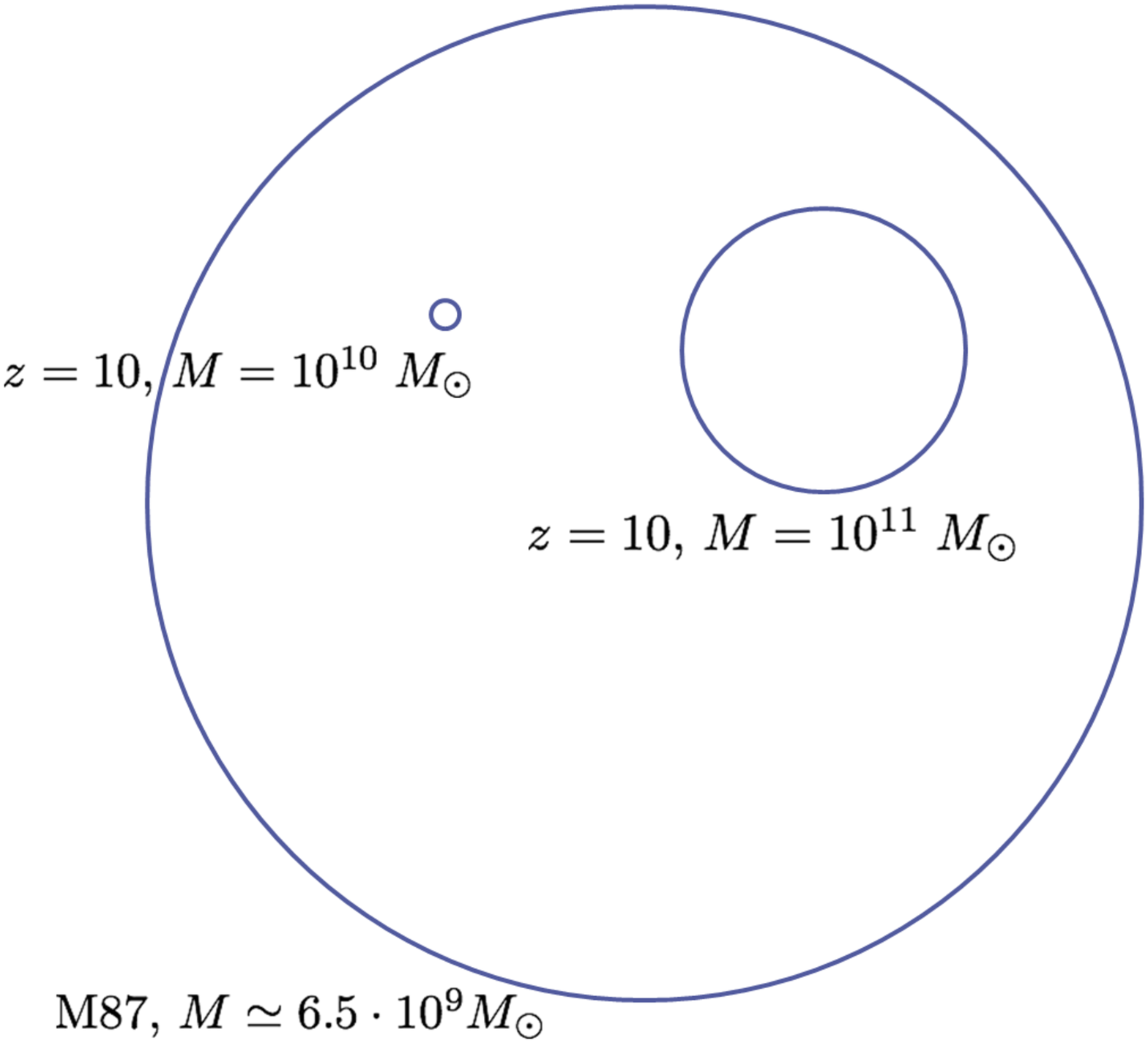}
\bigskip

	\caption{Relative shadow size of BH shadow for different masses at different redshifts. Image by O. Tsupko. \label{fig18}
}}
	    \end{figure}

The relative size of shadows of different mass BH, at redshift $z=10$ are presented in Fig.\ref{fig18}. Note also the recent paper  \cite{Mehrgan-2019} where the discovery of supermassive black hole with the mass of $(4.0 \pm 0.80) \times 10^{10} M_{\odot}$  in the galaxy Holm 15A, at redshift of  z = 0.055, is reported. This is an indication that the masses $\sim \, 10^{11} M_\odot$ may be also expected.

\section{Conclusions}

1. The Schwarzschild black hole shadow was studied analytically for local sources, and with account of a cosmic expansion.

2. The exact analytical solution is found for the expansion driven by a cosmological constant only. It is found that the angular radius of the shadow shrinks to a nonzero finite value if the comoving observer approaches infinity.

3. For a general case of cosmic expansion (with matter, radiation and Lambda) the approximate solution is found for dependence of  the angular size of the shadow on the redshift.

4. The shadow size of supermassive BHs at cosmological distances increases with redshift in presence of the matter component, and may reach the values comparable to the shadow size in M87, and in the center of our Galaxy.

\section*{Aknowledgements}

GSBK and OYuT acknowledge financial support by Russian Science Foundation, Grant No. 18-12-00378.

\bigskip
\bigskip
\noindent {\bf DISCUSSION}

\bigskip
\noindent {\bf D. BISIKALO:} Could you explain observable asymmetries in the in the shadow and the ring around?

\bigskip
\noindent {\bf G.S. BISNOVATYI-KOGAN:}
The visible asymmetry may be connected with doppler and boosting effects of the accretion disk with low angle inclination to the plane of the sight.

\bigskip
\noindent {\bf WOLFGANG KUNDT's Comment:} No shadow without a lamp. The $1.3$ mm line seen towards M87 implies a lamp of
T = $10^{9.5}$ K , which cannot be emitted by a SMBH - see my talk to-morrow - but
has been interpreted by me as a BD since 1978.

\end{document}